# Observations of Ultra-High Energy Cosmic Rays

**A A Watson**

School of Physics and Astronomy, University of Leeds, Leeds, LS2 9JT, UK

a.a.watson@leeds.ac.uk

**Abstract**. The status of measurements of the arrival directions, mass composition and energy spectrum of cosmic rays above $3 \times 10^{18}$ eV (3 EeV) is reviewed using reports presented at the 29th International Cosmic Ray Conference held in Pune, India, in August 2005.

## 1. Introduction

There is considerable uncertainty about the properties of the highest energy cosmic rays. The major questions concerning possible anisotropies in the arrival direction distribution, the mass composition and the long-sought steepening in the energy spectrum near $10^{20}$ eV (the Greisen-Zatsepin-Kuzmin effect) remain unanswered, but there has been significant progress. In this article I describe the status of each of these topics, as reported during the recent International Cosmic Ray Conference held in Pune, India, in August 2005 (ICRC2005), focussing on data above $3 \times 10^{18}$ eV (3 EeV). This was the first occasion at which results from the Pierre Auger Observatory, now more than 50% complete, were presented in detail and accordingly some discussion of the performance of this instrument is given. Unfortunately, none of the major questions can yet be answered with confidence: accordingly discussion of interpretations of the results is very restricted.

## 2. The Exposure and Event numbers from various Instruments

In Table 1 the approximate exposure and the numbers of events claimed to be above 3 and 10 EeV from instruments currently or recently operating is shown. The aperture assumed in calculating each exposure is appropriate to ~ 10 EeV. In all cases, except that of the Auger Observatory, these are functions of energy above 3 EeV, although for AGASA and Yakutsk the apertures saturate above ~ 10 EeV. The AGASA Observatory has now closed, while the HiRes instruments will be operated until March 2006. The Yakutsk array, now relatively small by modern standards, continues to take data. Construction of the Telescope Array, led by a Japanese consortium in Millard County, Utah, USA, has begun and a small number of scintillators are operating, overlooked by a fluorescence detector. A cursory comparison of the rate of events above 10 EeV makes clear that the differences between the integral rates are much larger (by more than a factor of 2) than can be accounted for by Poissonian variations. Possible reasons for the discrepancies will be discussed below.

## 3. The Pierre Auger Observatory

During ICRC2005, the Pierre Auger Collaboration reported their first results on the energy spectrum and arrival direction distribution from the Observatory. This new device has been designed to combine the best features of surface detectors (SD) and fluorescence detectors (FD) in a 'hybrid' instrument and there is promise of definitive measurements from it within the next few years.

Table 1: Exposure and Event Numbers from various instruments

|  | km$^2$ sr yr | > 3EeV | >10 EeV | rate > 10 EeV km$^{-2}$ sr$^{-1}$ yr$^{-1}$ |
|---|---|---|---|---|
| **AGASA** | 1600 | 7000 | 827 | 0.52 |
| **HiResI mono** | ~5000 | 1616 | 403 | 0.08 |
| **HiResII mono** |  | 670 | 95 |  |
| **HiRes Stereo** | ~2500 | ~3000 | ~500 | ~0.20 |
| **Yakutsk** | ~900 | 1303 | 171 | 0.19 |
| **Auger** | 1750 | 3525 | 444 | 0.25 |

**Telescope Array**: The plan is for 760 km$^2$ with 576 scintillators and 3 fluorescence detectors.

Construction of the prototype for the Auger Observatory began shortly after a ground-breaking ceremony in March 1999. The performance of the prototype has been described [1] and only relatively minor changes were made to the design of the sub-systems of the instrument now under construction. A data run began in January 2004 and continued into June 2005. At the start ~150 water-Cherenkov detectors, each of 10 m$^2$ and 1500 m from each other, were operational, together with one complete and one partially-complete fluorescence detectors. At the end of the run, the number of operational tanks had increased to ~780 and three of the final four fluorescence detectors were fully-functional. It has proved to be a relatively straight-forward matter to take high-quality data as the area grows. Completion of the Observatory is expected in early 2007. A large number of 'stereo events' has been recorded in which two fluorescence detectors recorded signals in coincidence with information from the surface detectors. In August 2005 the first tri-ocular event was detected. The status of the Observatory was described at ICRC2005 [2].

The high-level of understanding that is derived from being able to make simultaneous observations of the fluorescence signals and the water-tank signals is well-illustrated by results from the detection of the scattered light from the Central Laser Facility [3]. This facility, located close to the centre of the array, hosts a 355 nm frequency-tripled YAG laser that generates pulses of up to ~7 mJ. The scattered light seen at the fluorescence detectors from such a pulse is comparable to what is expected from a shower initiated by a primary of $10^{20}$ eV: the laser can be pointed in any direction. Some of the light from it is fed into an adjacent tank via an optical fibre so that correlated timing signals can be registered. In this way it has been established that the angular resolution of the surface detectors is ~ 1.7° for 3 < E < 10 EeV and ~ 0.6° for hybrid events. It has been shown [4] that the accuracy of reconstruction of the position of the laser using the hybrid technique is < 60 m. The corresponding figure for the root mean square spread is ~ 570 m if a monocular reconstruction is made. As there is always at least one tank response in coincidence with every detection by a fluorescence station, these data give a preliminary indication of the geometrical power of the hybrid technique.

The ability to make simultaneous observations of air-showers with the SD array and the FD is a unique feature of the Auger Observatory and has been used to calibrate the measurements made with the SD so that a primary energy can be estimated, relatively independently of assumptions about the mass of the primary particle and with a relatively small systematic uncertainty arising from lack of knowledge of hadronic interactions. A study of the 20 highest energy events has demonstrated that the properties of these showers are in no way unexpected when compared with the more numerous lower energy events that have been recorded so far: a large event (E > 140 EeV) was detected but fell with its core outside the fiducial area as then defined [5].

**4. The Mass Composition of Cosmic Rays**

Important developments in our understanding of hadronic interaction models at the highest energies were reported at ICRC2005. In particular a revision of the QGSJET series of models (QGSJETII) was presented by Ostapchenko and Heck [6]. Their new analysis of high-energy interactions has led to improved parton distribution functions that are consistent with cross-section measurements. It is found that there is a reduction in the number of secondary particles produced in hadron-nucleus and nucleus-nucleus collisions leading to an increase in the number of electrons observed at ground level. Of particular importance for mass inferences at the highest energy are the results on the depth of shower maximum, $X_{max}$, and the number of muons. $X_{max}$ is now predicted to be some 10% deeper in the atmosphere for both proton and iron primaries than was the case with QGSJET01, while the muon signal is 83% of that found formerly, assuming proton primaries, and 90% of what had been expected for iron nuclei at 10 EeV.

Qualitatively, the behaviour of $X_{max}$ and muon number with energy calculated with the new model is similar to what has been found using Sibyll 2.1 [7]. It is clear that significant reinterpretations of existing data on the muon component and on $X_{max}$ are called for. Also, conclusions previously drawn from observed fluctuations in $X_{max}$ will need to be reassessed. It has been pointed out previously [8] that the mean mass above 10 EeV might not be as proton-rich as has often been assumed and the new model results appear to confirm this hypothesis. While it is almost certain that there will be more surprises as accelerator measurements are made at higher energies, the possibility of a higher mean mass than has been hitherto assumed needs to be kept in mind when interpreting data on the energy spectrum and arrival direction pattern of ultra-high energy cosmic rays.

**5. The cosmic ray energy spectrum above 3 EeV**
Recent measurements of the high-energy cosmic ray spectrum by the AGASA [9] and HiRes [10] collaborations have yielded conflicting results (see table 1). This may be because there are serious limitations in using a surface detector (SD) array or a fluorescence detector (FD) system independently to measure the primary spectrum. In the former case, observations at ground level are used to derive the energy of the primary particle and assumptions about the nature of the primary and of features of hadronic interactions at energies well-above accelerator measurements are unavoidable. In the latter case, as with any fluorescence instrument, the monitoring of the atmosphere is a major problem and additionally, there are significant uncertainties in determination of the aperture of the detector for which assumptions about the mass of the primary and of the energy spectrum are necessary. The method developed by the Auger Collaboration largely circumvents these problems and leads to a reliable estimate of the primary energy spectrum. The Collaboration has reported [11] the first precision measurement of the high-energy cosmic ray spectrum made from the Southern Hemisphere.

For this analysis attention was restricted to events with zenith angle, θ, <60°. The strategy is to reconstruct the arrival direction for each event recorded by the SD and to estimate the magnitude of the signal at 1 km from the shower axis, S(1000), as a measure of the size of the shower in units defined by the signal from a muon that traverses the tank vertically. S(1000) is chosen as the ground-parameter as it can be measured to better than 10%. In addition, as shown in the pioneering studies of Hillas [12], the size of this ground-parameter is ~ 3 times less susceptible to stochastic fluctuations and variations in primary mass than are measurements made close to the shower axis.

Two cosmic rays of the same energy, but incident at different zenith angles, will yield different values of S(1000). Thus a necessary step is to find the relation between the ground-parameter measured at one zenith angle and that measured at another. The approach adopted here is to use the well-established technique of the constant intensity cut (CIC) method which has been recently reappraised [13]. The principle of this method is that the high level of isotropy of cosmic rays leads to the proposition that showers created by primaries of the same mass and energy will detected at the observation level at the same rate. Here the rate of events above different S(1000) is found for different zenith angles and all azimuth angles so that events come from a broad band of sky. This method is used to establish the relationship between $S(1000)_{38°}$ and $S(1000)_\theta$, where the subscripts refer to a reference angle, chosen as 38°, and θ is the angle of incidence. The average thickness of the atmosphere above the Auger Observatory is 875.5 g cm$^{-2}$.

The link between $S(1000)_{38°}$ and the primary energy is best established using data from the fluorescence detectors rather than through model calculations. On clear, moonless nights, it is possible to observe fluorescence signals simultaneously with the SD events: this 'hybrid' approach, a key characteristic of the Auger Observatory, offers several advantages. For every FD event for which the shower core falls within the instrumented SD area, at least one tank is struck so that the time at which the tank was triggered can be used to enhance the reconstruction of the FD geometry. Further, as the FD instruments are used primarily as calibration devices in this application, the selection of events can be made in a highly selective manner. This was done in [11], where the FD tracks had to be longer than 350 g cm$^{-2}$, the contribution of the Cherenkov light to the signals collected less than 10% and there were contemporaneous measurements of the aerosol content of the atmosphere, as was possible in the latter part of the data run. There are significant systematic uncertainties currently present in the Auger spectrum arising largely from the lack of knowledge of the fluorescence yield of atmospheric nitrogen and from the low statistics available for the $S(1000)_{38°}$ energy calibration. At 3 EeV the systematic uncertainty is about 30% growing to 50% at 100 EeV.

When estimating the energy of an event from the fluorescence yield, a correction must be made for 'missing energy' carried by high-energy muons and neutrinos. A study of this conversion factor has recently been made for nucleonic primaries with a variety of hadronic interaction models. At 10 EeV the correction for missing energy is ~10% with a systematic uncertainty, due to our lack of knowledge of the nuclear mass and the hadronic interactions, estimated as ~7% [14]. The corrections and the associated systematic uncertainties may have to be revised when LHC data are available. In deriving these figures, a nuclear mass between proton and iron has been assumed. If the high-energy beam contained a mixture of photons and nuclei, use of the average energy calibration relation derived from Auger Observations would lead to an under-estimate of the energy of any nuclei in the primary beam and an over-estimate of the energy of any photons that are present. This is because a photon shower yields an S(1000) that is ~ 4 smaller than that from a proton or Fe-nucleus of the same energy. The magnitudes of the under- or over-estimates would depend on the fraction of photons present. At present, the upper limits to the fraction of photons, at 95% confidence levels, are 26% above 10 EeV [15], 50% above 40 EeV [16] and 70% at $10^{20}$ eV [17].

For a surface array, the primary energy is derived under particular assumptions about the mass of the primary particle and about the hadronic interactions. A suitable conversion relation has yet to be derived using QGSJETII and the mass is clearly uncertain as inferred from the $X_{max}$ measurements [6]. For example, if the energies of the AGASA events near 100 EeV are estimated using protons and the SIBYLL model or iron with the QGSJET01 model, the ratio of the energies is 1.33.

The exposure of the Auger Observatory has had to be determined with some care as the number of operating tanks of the SD has increased from ~150 in January 2004 to ~780 in June 2005. It was computed as 1750 km$^2$ sr yr and is independent of energy above 3 EeV. This exposure is slightly above that achieved by the AGASA group [9] and is a factor of about 3 lower than the exposure of the monocular HiResI [10]. The HiRes stereo exposure, from which the most reliable fluorescence-only energy spectrum is expected, is not yet available. The average array size during the time of this exposure was 22% of what will be available when the Auger southern site has been completed.

For the HiRes detectors (I and II and stereo) the aperture changes with energy and is never energy independent. For example, the stereo aperture grows from 6 x $10^2$ km$^2$ sr at 3 EeV to 1.5 x $10^3$ km$^2$ sr at 10 EeV. The problems in determining the apertures, which depend on assumptions about high energy interactions, and about the mass and spectrum of the primaries are discussed in papers at the ICRC2005 [18]. While understanding of the problem is growing, it is evident that the situation is not yet finalised. Furthermore, the aperture depends on a detailed understanding of the atmospheric conditions, which need to be known on a shower-by-shower basis. The spectra so far published by the HiRes team assume an average atmosphere using data on a monthly basis, rather than measurements on a much shorter timescale. It is possible for example, that the variation of aperture with energy is more rapid than is claimed, in which case the spectrum slope would be steeper and claims for the GZK-steepening [10] would be less firm. No stereo spectrum has yet been reported.

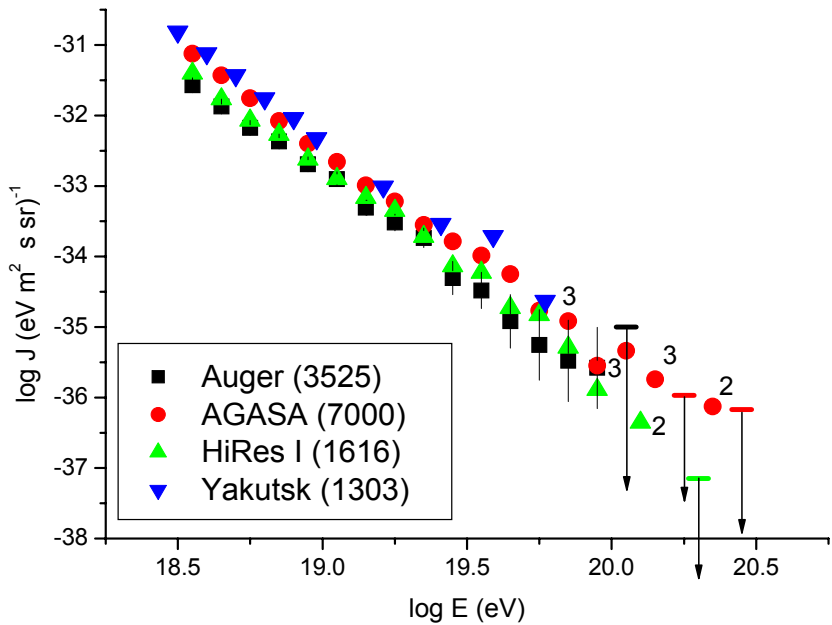

**Figure 1.** The differential spectra from Auger, AGASA, HiResI and Yakutsk are compared on a plot of log J vs. log E. The numbers shown in the legend correspond to the events reported above 3 EeV. The numbers (3, 2) by some points refer to the last bin of each data set in which > 0 events were recorded.

A comparison of spectra measured by Auger, AGASA, HiResI and Yakutsk is made in figure 1. In figure 2 the ratio of the values from Auger, AGASA and HiResI are compared with an $E^{-3}$ line passing through the Auger point at 3.55 EeV which contains 1216 events. It is clear that further statistics are needed and that a better understanding of the conversion to energy in the case of SD arrays is required.

The differences between the fluorescence measurements by Auger and HiResI are relatively small except at the highest energies where the Auger statistics are presently too low to comment on the flux above 100 EeV. The difference between AGASA and the fluorescence measurements probably arises, at least in part, because of the mass and hadronic model assumptions. If these differences become established, a promising route to extracting model and mass information will be open.

## 6. Studies of Arrival Directions at High Energies: is there evidence for clustering?

For some time the AGASA group [19] have reported clustering on an angular scale of 2.5°, from a data set of 59 events above 40 EeV. Such clusters are claimed to occur much more frequently than expected with an estimate of $10^{-4}$ given for the chance probability. A search of the HiRes data [20] has not revealed clusters with the same frequency as claimed by AGASA. Finley and Westerhoff [21] have presented an analysis using the directions of 72 events recently released by the AGASA group.

They have taken the 30 events described in [22] as the trial data set and used the additional 42 events to search for pairs, adopting the criteria established by the AGASA group. Two pairs were found: such a result is estimated as having a probability of 19% of occurring by chance.

A further search for clusters has been made by the HiRes group using 27 events from their own data and 57 from AGASA above 40 EeV [23]. Using a novel likelihood search method, the authors state that "no statistically significant clustering of events consistent with a point source is found": the most significant effect is the AGASA triplet. If the energy scales of the two instruments are then normalised by reducing the HiRes threshold to 30 EeV, the HiRes sample is increased to 40 events.

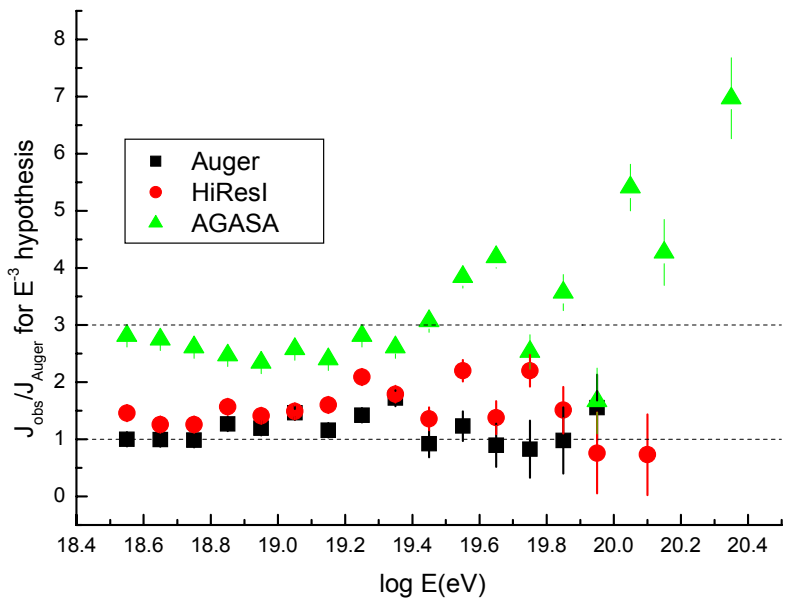

**Figure 2.** The ratio of the values of each point with respect to a fit of $E^{-3}$ to the first point of the Auger spectrum at 3.55 EeV which contains 1216 events. The purpose of the plot is to illustrate the differences between the different measurements in a straight-forward manner. Yakutsk data are not included in this plot as they are so discordant (see figure 1).

An event close to the AGASA triplet is found in the resulting sample but, as the HiRes group make very clear, it is not possible to evaluate a valid chance probability for this observation because of the *ad hoc* nature of the energy shift. Accordingly, they have identified a direction (near $\alpha = 169°$ and $\delta = 57°$) and state that if 2 events from the next 40 observed with the same energy selection fall in a bin of $1°$ around this direction the chance probability will be $10^{-5}$. The target of 40 events should be reached by the time the HiRes Observatory is decommissioned in March 2006.

A feature of the HiRes instrument is the excellent angular resolution achieved with stereo events: a resolution of ~ $0.5°$ is claimed above 10 EeV. This is significantly better than was reached at AGASA and surpasses what has so far been demonstrated with the surface detectors of the Auger Observatory. Such resolution is particularly valuable in searches for small-scale anisotropy. At the Pune meeting, the group reported [24] the results of a search for correlations with BL Lac active galaxies. Claims of associations have been made previously based on data from AGASA and Yakutsk [25] but are regarded as controversial because of the way in which *a postiori* selections were made. In [26] an analysis of correlations of BL Lacs from the Veron 10th catalogue with m < 18 was reported using 271 HiRes events above 10 EeV. The HiRes group [24] have confirmed these claims and, speculating that the correlations are indicative of neutral particles, have extended the search to lower energies. For all 4495 HiRes stereo events, they find a significance $F = 2 \times 10^{-4}$, where F is the fraction of simulated event sets that have a likelihood of clustering by chance greater than is found in the data. The search has been extended to high polarisation BL Lacs and also to those BL Lacs that are TeV γ-ray sources.

Small values of F, corresponding to probabilities less than 0.5%, are found in several cases and these selections are now regarded as hypotheses to be tested with additional HiRes data.

## 7. Summary
When an energy spectrum based on fluorescence detectors has been finalised and compared with one from surface detectors in which mass and hadronic model assumptions are of key importance (such as that from AGASA) conclusions about the issues of mass and models are to be expected. One of the most important results from ICRC2005 is the new insight that comes from the use of QGSJETII as the hadronic interaction model. A consequence is that the mean primary mass above 3 EeV may be higher than thought hitherto. This has implications for reconciling the SD and FD spectra, for explaining the isotropy of the highest energy cosmic rays and for easing the problem of UHECR acceleration.

**Acknowledgements**
I would like to thank the organisers of the excellent TAUP meeting in Zaragoza for the invitation to give a review talk and my colleagues within the Auger Collaboration for their constant stimulation. Research on Ultra High Energy Cosmic Rays at the University of Leeds is supported by PPARC, UK.